\documentclass[AMA,STIX1COL]{WileyNJD-v2}

\articletype{Article Type}%

\received{<day> <Month>, <year>}
\revised{<day> <Month>, <year>}
\accepted{<day> <Month>, <year>}

\begin{document}

\title{A survey of security and privacy issues in the Internet of Things from
the layered context}

\author[1]{Samundra Deep}

\author[1]{Xi Zheng*}

\author[1]{Alireza Jolfaei}

\author[2]{Dongjin Yu}

\author[3]{Pouya Ostovari}

\author[4]{Ali Kashif Bashir}

\authormark{SAMUNDRA \textsc{et al}}

\address[1]{\orgdiv{Department of Computing}, \orgname{Macquarie University}, \orgaddress{\state{Sydney, NSW}, \country{Australia}}}

\address[2]{\orgdiv{College of Computer Science and Technology}, \orgname{Hangzhou Dianzi University}, \orgaddress{\state{Hangzhou}, \country{China}}}

\address[3]{\orgdiv{Charles W. Davidson College of Engineering}, \orgname{San Jose State University}, \country{USA}}

\address[4]{\orgdiv{Department of Computing and Mathematics}, \orgname{Manchester Metropolitan University}, \country{UK}}

\corres{*Xi Zheng, Macquarie University. \email{james.zheng@mq.edu.au}}


\abstract[Summary]{Internet of Things (IoT) is a novel paradigm, which not only facilitates a large number of devices to be ubiquitously connected over the Internet but also provides a mechanism to remotely control these devices. The IoT is pervasive and is almost an integral part of our daily life. These connected devices often obtain user's personal data and store it online. The security of collected data is a big concern in recent time. As devices are becoming increasingly connected, privacy and security issues become more and more critical and these need to be addressed on an urgent basis. IoT implementations and devices are eminently prone to threats that could compromise the security and privacy of the consumers, which, in turn, could influence its practical deployment. In recent past, some research has been carried out to secure IoT devices with an intention to alleviate the security concerns of users. There have been research on blockchain technologies to tackle the privacy and security issues of the collected data in IoT. The purpose of this paper is to highlight the security and privacy issues in IoT systems. To this effect, the paper examines the security issues at each layer in the IoT protocol stack, identifies the under-lying challenges and key security requirements and provides a brief overview of existing security solutions to safeguard the IoT from the layered context.}


\maketitle


%
%
\section{Introduction}\label{sec1}

The term Internet of Things (IoT) was coined by Kevin Ashton in 1998 \cite{serbanati2011building}. The concept of IoT was presented as an idea to link RFID tags to the Internet. The general definition for the IoT includes heterogeneous devices and the interconnection of these uniquely identifiable objects. The IoT is a network of several internet-connected physical devices. These interconnected devices are embedded with various sensors which make them smart enough to gather data and exchange information with other devices in the network. This would enable any device to communicate with any other device thereby creating a smart ecosystem. IoT is a dominant technological revolution that refresh the present Internet foundation to an idea of substantially more propelled computing system in which various smart devices are pervasively connected and uniquely addressed  \cite{singh2014survey}.

IoT is a collection of heterogeneous technologies that function together. IoT devices are equipped with various components such as actuators, embedded sensors, processors, RFID and transceivers for acuity, concurrence and connection. The core purpose of the IoT is to enable heterogeneous devices connect to the Internet and exchange information among inter-connected devices in a reliable manner \cite{khan2012future}. The security of network and protection of data should meet essential standards and basic principles of integrity, authentication, availability, authorization and confidentiality of user's information \cite{atzori2010internet}. Human life  has been enriched with the availability of smart and intelligent IoT devices in hospitals, homes, transportation systems and aged care facilities, etc.\cite{zhao2013survey,zhang2019driver,bhandari2017non,abkenar2017service, lu2019detection}. IoT unquestionably has a huge potential for adaptability and guarantees an extraordinary future. However, it is inherently prone to threats that can compromise the security and privacy of the users, which could in turn influence its practical development~\cite{yu2018survey,pan2017cyber,zheng2017security,radhappa2018practical,zheng2017testbed,zhang2018soprotector,zheng2016investigating}. Eventually confidentiality, integrity and availability of the data will be diminished and as a result clients will be hesitant to accept this technology \cite{kanuparthi2013hardware}. 

\begin{figure}[ht]
\centerline{\includegraphics[width= 3.8 in]{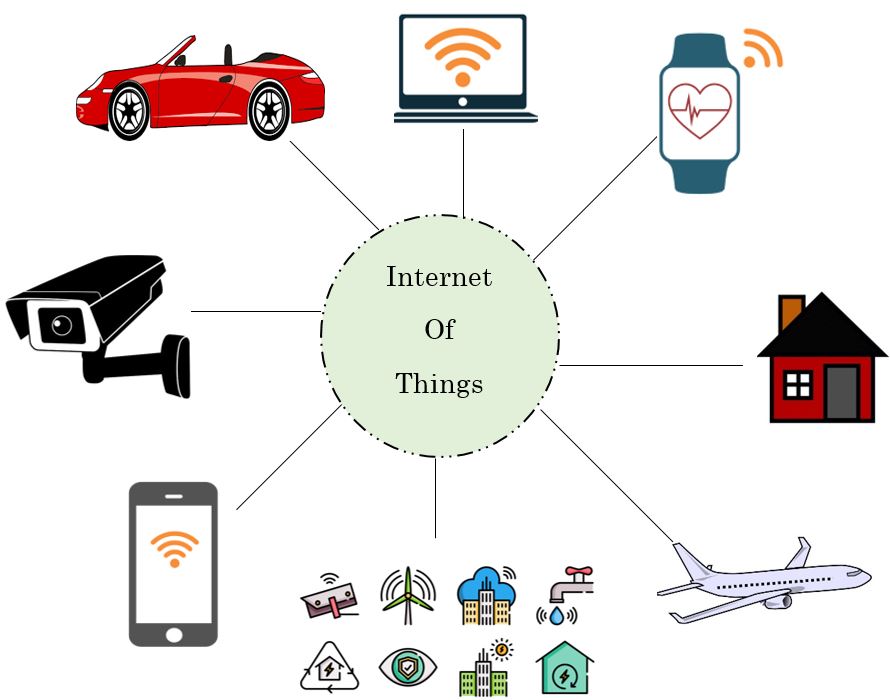}}
\caption{Internet of Things Applications \label{fig1}}
\end{figure} 

The large number of devices which are being connected to the Internet is rapidly proliferating and this may eventually lead to an all pervasive IoT enabled global Internet architecture. A study by Gartner estimates around 25 billion addressable smart devices are likely to be connected to the computing network by 2020 \cite{Gartner} with a decent number of these being appliances, therefore there will be an expansive open door for the hackers to utilize these gadgets to their own advantage through malicious emails, ``denial of service'' attacks and by various other malicious means such as unsafe worms or Trojans. Thus, security concerns are noteworthy part that has to be well studied before developing more advanced Internet of Things (IoT) systems \cite{ali2018quality}. Billions of data collected and stored online are not secured. There have been researches on potential of blockchain technology to secure the online data generated every seconds \cite{panarello2018blockchain}. Blockchain technology is considered innately secure since it uses block of records of transactions. All the blocks are inter-connected, consequently, it is difficult for the hacker to tamper the data with just a single record \cite{khan2018iot}. It also use powerful cryptography to secure the chain of the data.

Our survey paper reviews and analyses the security and privacy issues in IoT. The rest of the paper is as follows: Section 2 contains the background of IoT. 
In Section 3, we examine the security issues at each layer of the IoT protocol stack and identify corresponding threats and attacks that can manifest at each layer. In Section 4, we address the major challenges in securing IoT and discuss the security services requirements in IoT. In Section 5, we provide an overview of existing approaches to securing the IoT. We discuss some of the research directions in Section 6. Finally, in Section 7 we provide some concluding remarks.

\section{Background
}\label{sec2}

\begin{figure}[ht]
\centerline{\includegraphics[width=\linewidth]{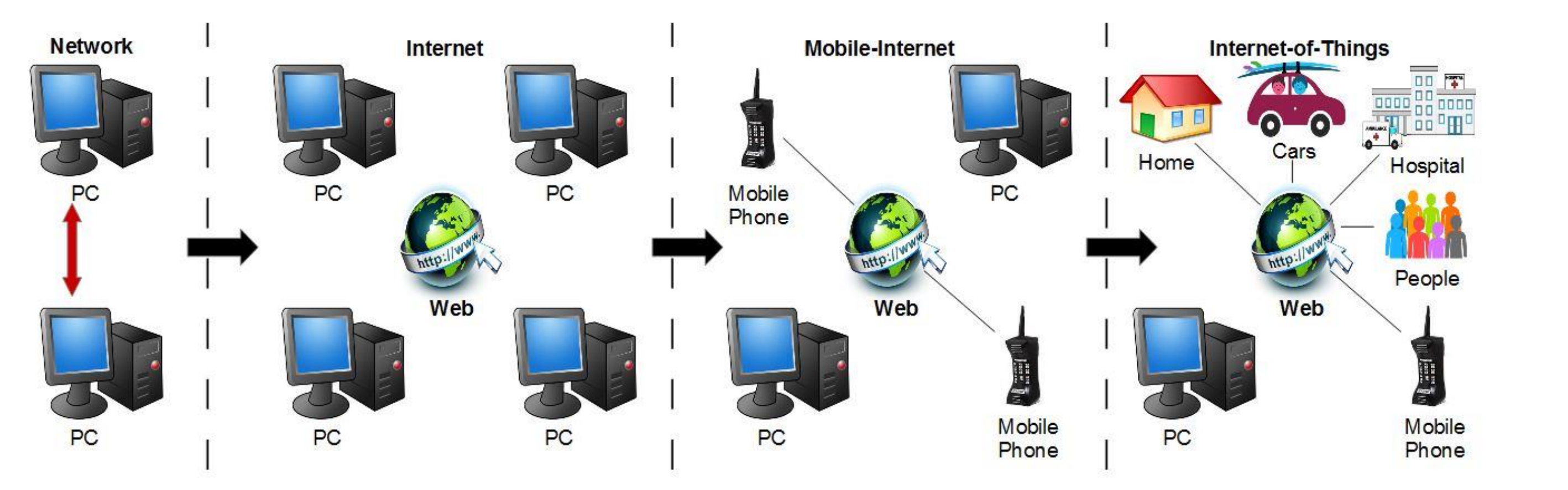}}
\caption{Evolution of the Internet of Things\label{evolu}}
\end{figure}

IoT is believed to be the most influential technology of the generation after Internet. The number of interconnected physical devices is significantly increasing and it has already surpassed the human population in 2010. There has been significant work in the development of IoT-enabled devices in the recent time. The advances in the technologies in terms of resources constrained and energy-efficient devices have extended the outreach of internet even to the remote locations also \cite{musaddiq2018survey, kraijak2015survey}. The number of interconnected physical devices have exceeded all expectations. The evolution of the Internet of Things is as shown in Figure \ref{evolu}.

\begin{figure}[ht]
\centerline{\includegraphics[width=\linewidth]{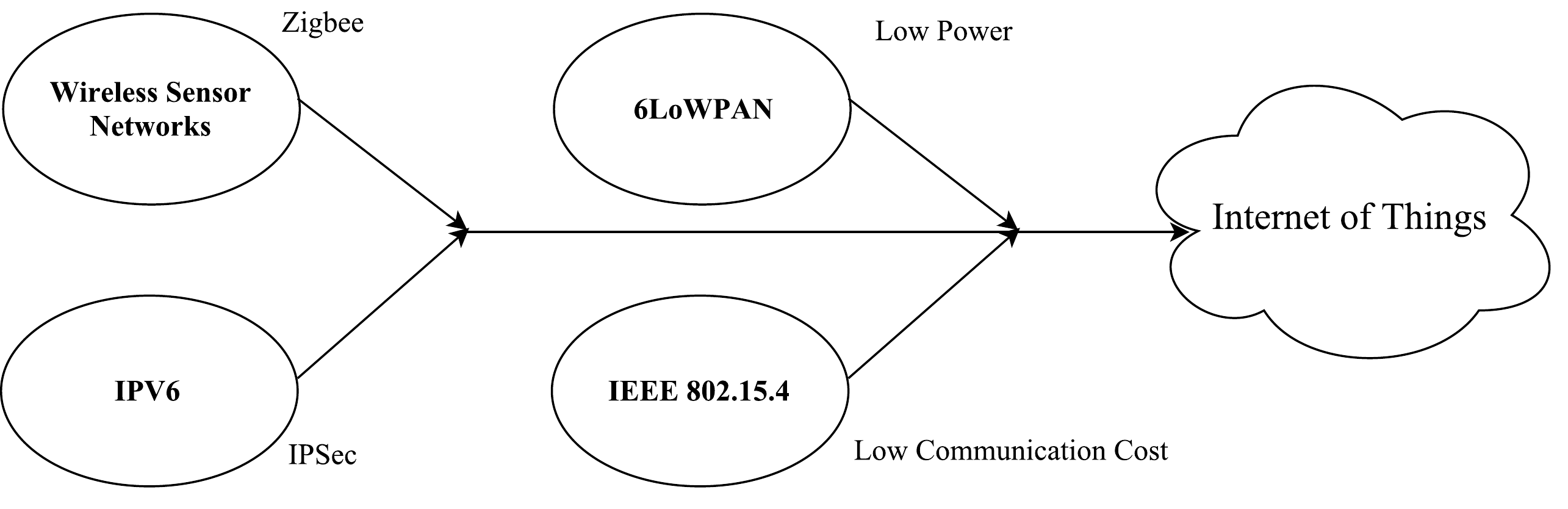}}
\caption{Background of the Internet of Things \label{background-iot}}
\end{figure}

In 1996, the Internet Engineering Task Force (IETF) defined IPv6 addressing and the advancement in this area has propelled the evolution of the IoT devices \cite{vaughan2003mobile}. The technologies such as IEEE 802.15.4, 6LoWPAN, IPv6 are all defined to support the necessities of the present day internet. The background of the Internet of Things are as shown in Figure \ref{background-iot}. The IoT is not only about the smart devices such as computers, smart phones, tablets that are connected to the Internet. IoT is heterogeneous technology in which all the small and simple devices around us can be connected to the network and are able to communicate with each other. Advances in the area of electronics and communications have propelled internet expansion from PCs, smart devices etc to other physical devices. The technologies such as Bluetooth and WiFi have extended the network capabilities of devices. The evolving advancements in the VLSI design technologies with the progress of time have cut down the cost and sizes of the devices \cite{kumar2018review}. The progress in the technology has made devices and sensors considerably small so that it can be deployed broadly for various purposes and can be managed easily. The improvement over communication systems enable these devices and sensors to communicate with each other in the network and allow to share lot of information. The sensor technology plays an important part in the advancement of the IoT since it is the direct medium of gathering information from individuals.

These devices have lower power requirements however they are constrained due to their low memory capacity and processing capabilities. With regard to wireless networking, 6LoWPAN was introduced which is essential for low powered networking environments \cite{montenegro2007transmission}. The 6LoWPAN \cite{lin2017survey} has evolved as an essential element of the IoT system at this point. The issues identified with IoT technologies have attracted many professional and industrial interests in recent years. Several organizations and institutions are engaged in the advancement of the Internet of Things. IEEE is clearly the most important agency that is working for the standardization of IoT. IEEE project P2413 involves the work related to standard architectural framework for IoT \cite{ieee2016p2413}. The EU Commission officially announced The Alliance for Internet of Things Innovation in 2015 which is engaged in the research for the development of IoT technologies.


A standard architecture design for IoT is still an open issue \cite{jing2014security}. In order to realize the full advantages offered by the IoT, a standard model for communication between the participating devices is necessary. Several international organizations such as International Telecommunication Union, IEEE etc are actively engaged for the development and standardization of IoT \cite{internet2016international}. However, basic ideas are already defined that can act as base for the development of the Internet of Things. The technologies such as IEEE 802.15.4, 6LoWPAN, IPv6 etc. are all defined to support the necessities of the future internet. Currently, TCP/IP protocol stack is widely used in the Internet for data exchange among network hosts \cite{khan2012future}, \cite{silva2018internet}. The IoT is a large domain that incorporates an extremely wide range of technologies, from constrained to unconstrained devices and from stateful to stateless devices. Therefore, on which architecture paradigm to use is in place can be used as an outline for all possible implementations. For instance, the IoT architectures based on sensors, actuators or RFID tags are different from each other. A reference model can possibly be recognized however, it is probable that several reference architectures will co-exist in the IoT domain.

\begin{figure}[h]
\centerline{\includegraphics[width= 3.2 in]{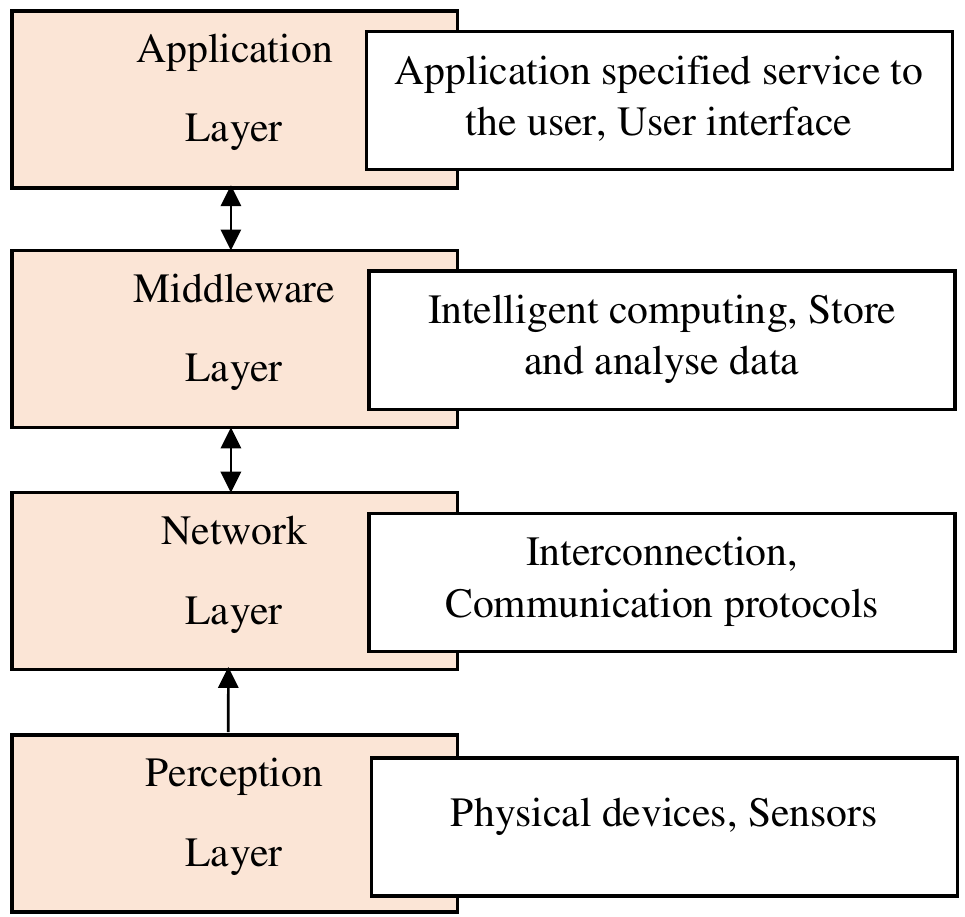}}
\caption{General IoT Architecture}
\label{genral-arch-iot}
\end{figure}

There are numerous architecture proposed for IoT by different researchers and most of them focus on middleware layer and network layer that deal with the necessities of IoT. The basic architecture proposed for IoT are as shown in Figure \ref{genral-arch-iot} namely, perception layer, network layer, middleware layer and application layer \cite{adat2017security}. These layers are responsible for information exchange in the network and provide application specified service to the user. The perception layer is responsible for sensing and collection of data through sensor nodes and other hardware \cite{puthal2016threats}. The network layer enables connectivity with other smart devices and the internet. It is used to transmit and process sensor data \cite{bello2017network}. The middleware layer is proposed to be between the network layer and the application layer. This layer is responsible for making intelligent decisions based on the processed results. It provides efficient delivery of services assuring scalability and interoperability \cite{da2018reference}. Application layer supports business services and is responsible for analysing the received information and making intelligent decisions to meet the requirements of the users, such as when to perform what things \cite{jing2014security}. The application layer comprises of different applications for the business needs for example, Constrained Application Protocol (CoAP) which is a replacement for the HTTP for resource-constrained devices \cite{saadeh2016authentication}.

\section{Security Issues In Layered IoT Architecture}\label{sec3}
There have been numerous research and achievements in context of the IoT. However, there are open issues and challenges which still need to be addressed in securing the IoT \cite{raja2018internet}. Thus, the IoT systems are compromised with security issues and are vulnerable to many attacks. In this section, we examine the security issues at each layer of the IoT protocol stack and identify corresponding threats and attacks that can manifest at each layer. Some of the security issues/attacks in the layered IoT architecture are classified in the Table \ref{list}.

\begin{center}
\begin{table}[h]%

\centering
\caption{list of security challenges, issues/attacks in existing IoT systems\label{list}}
\begin{tabular}{p{2cm} p{9.8cm} p{3.3cm} }  
\toprule
\textbf{IoT \newline Layers} & \textbf{Security Issues/ attacks}   & \textbf{Security  Parameters} \\
\midrule
Application Layer & Data access and security authentication issues, Data Protection and Recovery problems, spear-phishing attack, Software Vulnerabilities,  Attacks on Reliability and Clone attack \cite{kumar2016security} \newline & Data privacy, \newline Access Control \\ 

Middleware Layer & making intelligent decision processing huge data, malicious-code attacks, multi-party authentication, handling suspicious information \cite{razouk2017new} \newline & Integrity, \newline Confidentiality \\ 

Network Layer & Cluster security problems, DoS attacks, Spoofed, Altered or Replayed routing Information \cite{kraijak2015survey} \cite{turkanovic2014novel} \cite{padmavathi2009survey} \cite{padhy2011cloud} \newline & Authentication, \newline Integrity \\ 

Perception Layer & Node capture, Fake node, Mass node authentication, Cryptographic Algorithm and Key Management Mechanism \cite{zhao2013survey} & Integrity,\newline Authentication, \newline Confidentiality \\

\bottomrule
\end{tabular}
\end{table}
\end{center}

\subsection{Perception Layer}
The perception layer comprises of every single devices that are connected to a network in IoT and are in charge of exchanging information e.g., sensors, actuators, Zigbee, RFID frameworks, QR code and GPS systems \cite{jing2014security}. The security threats are at the node level in this layer. Most of the security issues arise from external entities such as embedded sensors, actuators etc. in the perception layer \cite{suo2012security}. Since the nodes are comprised of embedded sensors, actuators, transceivers etc., they become the prime targets of the attackers who aim to exploit them and try to substitute their own codes with the device software. Generally IoT devices lack battery power and memory capacity so they are simple, less powerful and light weight and are likely to  run out of power or affected by other external environment factors which could harm the functionality of  these devices, making it vulnerable to security attacks \cite{singh2017advanced}. Denial of service attack, fake node or malicious data, jamming, tampering, node capture etc. are most common attacks that occur in physical layer \cite{zhao2013survey}.

\subsection{Middleware Layer}
The middleware layer is an advanced layer above the network layer which does the mass data processing and make intelligent decisions \cite{bouloukakis2019automated}. It utilizes the advancements of technologies like cloud computing, big data processing and database. The layer has characteristics of processing massive amount of data hence it become difficult sometimes to manage huge data. The layer is capable of filtering valid and malicious data. However, to recognize valid data and filter out malicious information is a major issue in this layer \cite{kraijak2015survey}. The problem of handling suspicious information is another issue in this layer. The attacker can replace the data with the malicious information and can obtain lists of valid data and network informations. It can transmit the invalid or malicious information to the network that can lead to the failure or completely shut down of the network. Multi-party authentication for the resource constrained devices and securely data storing in the cloud are some of the prime concerns in this layer \cite{razouk2017new}.

\subsection{Network Layer}
Since the network layer carries a large amount of data it is highly susceptible to attacks which can lead to "network congestion" \cite{kumar2016security}. The core security issues in this layer are related to authentication and integrity of the data that are being transmitted in the network. Despite the relatively better security protection in the network layer, it is still vulnerable to counterfeit attack and Man-in-the-Middle attack \cite{kraijak2015survey}. The most common attacks that occur in the network layer are as follows:

\textbf{Replay attacks:} 
In replay attacks, an intruder copies a fragment or key of the messages that are being exchanged between the two parties and steals the information. The authenticated information is then re-transmitted maliciously to the receiver for an evil purpose, such as duplicate transactions \cite{turkanovic2014novel}. The authenticated message is sent again and again and the receiver processes the request believing it as a legitimate messages that meets the desire of the intruder.

\textbf{Denial of Service (DoS) attacks:}
Denial of Service (DoS) is a kind of attack in which the legitimate user is prevented from accessing the emails, websites, data or network services. It is a situation in which the attackers attempt to flood the network with unwanted traffic, malicious codes and actions that weakens the network's potential to provide the programmed service \cite{padmavathi2009survey}. 

\textbf{Man-in-the-Middle Attacks:}
In man-in-the-middle attack, an intruder intervenes in a communication between two parties, either to eavesdrop or to impersonate both parties and gains access to information of the communicating parties. The motive of the attack is to steal personal information, alter the message or data that are believed to be received from the trusted parties \cite{padhy2011cloud}. For instance, the recorded temperature by a sensor in IoT can be altered intentionally by the attacker to overheat the device, which lead to the failure of the working device.

\textbf{Malicious Code Injection:}
In malicious code injection, an attacker captures a working node and injects it with malicious-codes to gain access and control of the network. Sometimes it could also lead to the shutdown of the network \cite{fulare2011false}.

\textbf{Distributed Denial of Service (DDoS):}
A DDoS attack is a DoS attack in which a node is targeted by multiple compromised node by flooding the network with useless messages and malicious codes causing the unavailability of the service for the targeted users. The targeted networks are forced to slow down and often shut down leading to denial of service to the legitimate users.

\subsection{Application Layer}
The application layer includes smart devices that provide personalized services to the users. These devices are usually simple, low power and lightweight that are vulnerable to the attacks \cite{sethi2017internet}. Malicious attacks can replace the program codes with the bugs that may provoke the application to malfunction. Hence, the applications may compromised, shut down and fail to deliver what is programmed to do and also carry out authenticated services in inappropriate way. Application layer is responsible for data sharing, which can cause the problem of access control, privacy of data and leakage of information \cite{kraijak2015survey, jolfaei2019privacy}. Software vulnerabilities, spear-phishing attack, malicious code attacks, inability to receive security patches, hacking into the smart meter/grid are some of the common threats in the application layer \cite{kumar2016security}.

\section{Challenges And Security Requirements For IoT}

\subsection{Challenges in securing IoT}
Every new technology probably has some drawbacks and suffers from many challenges before and after they are deployed. Likewise, the Internet of Things also has some major challenges which make the user hesitant to accept this technology. Some of the challenges of IoT are discussed below: 

\subsubsection{Bandwidth and Power Consumption:} 
Generally IoT devices are designed to be lightweight, less powerful, less memory and small in size and they are not equipped with a large battery. IoT contains many interconnected devices and sensors to execute the programmed functionality with substantial security directions which may consume high bandwidth and drain out the devices. IoT systems should be well prepared with a concrete mechanism when there's any such unavailability of internet bandwidth. Hence, the minimization of bandwidth and power consumption remains a major challenges in the IoT. 

\subsubsection{Complexity:}
The Internet of Things consist of a network of internet-connected physical devices that have their own hardware/software layers and different system architectures for different purposes. All of these interconnected devices that are equipped with different sensors, actuators, protocols and standards accumulate together to execute the programmed function. Hence, it becomes more complex and hard to deal with this heterogeneous architecture in IoT systems \cite{talwana2016smart}.

\subsubsection{Sensing:} 
Internet of Things (IoT) consist of several internet-connected physical devices. IoT systems should be able to detect any loss of connectivity among the devices. The IoT system should be equipped with concrete sensing mechanism that could give the information about these devices, e.g., if there are any device failure inside the network or if the device has lost connectivity with the network \cite{talwana2016smart}. Monitoring of these several smart devices continuously and getting back them to the network again for connectivity after it had suffered from device failure or connectivity problems are also challenges for the Internet of Things.

\subsubsection{Lightweight Computing:} 
Since IoT devices normally have less memory capacity, traditional cryptographic-algorithms cannot be applied to the IoT system. Advanced cryptographic algorithms have high computing, storing and processing requirements which cannot be supported by IoT devices as they are resource constrained. Therefore, to find a way to implement required security mechanisms with low cost and minimum overhead is necessary for Internet of Things.

\subsection{Security requirements for IoT}
Security and privacy issues are the biggest concern for the IoT therefore, a detailed and accurate comprehension of security requirements in the aspect of IoT is crucial. Zhao et al. \cite{zhao2013survey} proposed some security requirements for secure IoT data transmission \cite{abbasi2019generalized}, which comprise: authentication and access control, appropriate secret key algorithms, physical security design, key management, secure routing protocols and intrusion detection technology. Weber et al. \cite{weber2015internet} have likewise proposed security requirements for secure IoT that include: access control, client Privacy, data authentication and attack resiliency. Vasilomanolakis et al. \cite{vasilomanolakis2015security} proposed security requirements that are categorized as: identity management, network security, resilience, privacy and trust. Earlier technologies like big data and cloud computing are probable to impart security services to the Internet of Things but the anomaly of the IoT acquaints additional problems to security which are not quite as same as the earlier technologies. Big data solutions for example are intended to scale and manage heterogeneity of information sources. The solution was not designed to manage constrained resources and uncontrolled environment \cite{vasilomanolakis2015security}. Similarly, Cloud computing are designed to scale and meet the requirements of the constrained resources and barely deals with the physical accessibility of sensors and mobility of devices \cite{vasilomanolakis2015security}. Some of the desired security service that are necessary for safeguarding the Internet of Things are discussed below:

\subsubsection{Confidentiality:} 
The confidentiality is a type of security services that prevent the unauthorized users from gaining access of the confidential data and information. It guarantees that the private information will not be diverted to the intruder and can only be gained by legitimate users. Different mechanisms and protocols have been proposed in the context of Internet of things to provide confidential service to the sensitive IoT data. Data Encryption, authentication and authorization process are some good practices to ensure confidentiality \cite{miorandi2012internet,jolfaei2015secure,jolfaei2012impact}.

\subsubsection{Availability:} 
The aim of the security services is to ensure the availability of data and services to the user, whenever required. Data availability is a security service that enables the user to access the information in the ordinary conditions as well as in the lamentable conditions. One of the serious threat to the data availability is DoS attack, which causes deny of service and make the data unavailable for the user \cite{bhabad2015internet}.

\subsubsection{Integrity:}
The integrity of the data is one of the important security services in IoT systems because the interconnected devices exchange sensitive information that could be altered or replaced by the attackers. The data integrity service ensures that the information which is being exchanged between the devices are originals and are not fabricated or modified by the hackers. There are different entities that can affect the originality of the data such as crashing of the server, failure of the sensor nodes etc. The data can also be modified by the intruders when it is being transmitted in the network. Read and write protection of data could be the solutions for such issues. Checksum and Cyclic Redundancy Check (CRC) are some basic error detection techniques for a segment of data to check the originality and accuracy of the data \cite{khan2012future}.

\section{Security Solutions For IoT}
Usually IoT devices communicate among themselves with negligible human interaction therefore mutual authentication is a critical aspect of the paradigm. The IoT devices are certainly designed for everyday usage and are generally used for gathering, storing and analyzing personal data. Thus, the devices should have the feature of being able to be controlled remotely and to support regular and automatic updates as well. Authentication and access control schemes should be applied in perception layer to prevent illegal node access and to identify network nodes \cite{saadeh2016authentication}. Data encryption and confidentiality schemes are extremely necessary to protect the collected data from modification and to prevent from malicious code injection \cite{jolfaei2015preserving, jolfaei2011substitution}. However, adding a strong data encryption and key management scheme would result in consuming resources of the IoT devices considerably. Hence, lightweight cryptographic algorithms and protocols are necessary to mitigate this problem \cite{jolfaei2017lightweight}. Intrusion detection systems can be implemented to detect any malicious behaviours in the network \cite{fulare2011false}. Several methods have been proposed in the past for securing the IoT devices and networks. We have listed security issues and attacks with a unique ID in Table \ref{ID}. We have concisely compared and analyzed the existing security solutions proposed in the recent literature which are discussed in Table \ref{perception}, \ref{network}, \ref{middle} and \ref{application}. 

\begin{center}
\begin{table}[h]

\centering
\caption{list of security challenges, issues/attacks in existing IoT systems\label{ID}}
\begin{tabular}{p{4.4cm} p{11.8cm} }  
\toprule
\textbf{Security attacks/threats unique ID} & \textbf{Security issues/ attacks}  \\
\midrule
C1 & Node capture, Fake node, Mass node authentication \\ 
	
C2 & threats involving the node security \\ 
	
C3 & RFID security problems \\ 
	
C4 & confidentiality service, key management \\
	
C5 & lightweight authentication and key establishment in wireless sensor networks (WSNs) \\ 

C6 & encryption and data integrity  \\ 

C7 & Distributed Denial of Service (DDoS) attackse  \\ 

C8 & masquerade attack, man-in-the-middle attack and replay attack \\

C9 & Data access and authentication issues, DoS attacks, secured communication session \\ 

C10 & peer authentication \\ 

C11 & lightweight cryptographic algorithms, store IoT data securely on the cloud database\\ 

C12 &  data authentication between the cloud and the smart devices\\

C13 & access control and authorization issues in interconnected devices\\ 

C14 & flexibility issues in authorization framework\\ 

C15 &  inter-device authentication issues and session-key distribution issues\\
C16 & security policy to address the privacy and security challenges\\ 

C17 & users privacy protection\\ 

\bottomrule
\end{tabular}
\end{table}
\end{center}

\subsection{Security solutions in the Perception Layer}
Perception layer comprises of various sensors devices,Radio Frequency IDentification RFID, Wireless Sensor Networks (WSNs),  GPS etc. Node capturing, attacks on the embedded sensors, Cryptographic Algorithm \& Key Management Mechanism are some prime targets of the attackers in this layer. Chen et al. \cite{chen2006distributed} proposed an algorithm to identify the compromised sensors in the wireless sensor networks. The author proposed the detection algorithm in which the sensors can identify their status in the distributed environment as "good" or "faulty". The status claimed by these sensors can be verified as accepted or rejected by the neighbouring nodes as they are capable of analysing the node behaviour. The experiment results showed that the proposed system has maximum accuracy and minimum false rate in detecting the faulty sensors in the WSNs. Also, the complexity of the algorithm is low. 
Li et al. \cite{li2013research} has pointed out the security issues in the perception layer and also some solutions to the issues are deliberated. The author has proposed an improvement on the PKI-like security mechanism protocol for the threats involving the node security in IoT systems. 
Aggarwal et al. \cite{aggarwal2012rfid} proposed an improved protocol for Radio Frequency Identification (RFID) security. The perception layers comprises of RFID tags in IoT system for gathering data and data communication between the connected devices. The author proposed an improved protocol that is computationally efficient and also prevent disclosure and desynchronization attacks. 

\begin{center}
\begin{table}[h]
\centering
\caption{Pros and cons of the addressed security challenges,issues/attacks in Perception Layer of existing IoT systems\label{perception}}%

\begin{tabular}{p{2cm} p{2cm} p{3.5cm} p{3.5cm} p{4cm} }  
\toprule
\textbf{Author and References} & \textbf{Addressed security solution Layers} & \textbf{Issues Addressed}& \textbf{Proposed Solutions}& \textbf{Pros and Cons of the existing solution} \\
\midrule
Chen et al. \cite{chen2006distributed} & Perception Layer & addressed the challenges mentioned in "C1" by identifing the compromised sensors in the wireless sensor networks & proposed a detection algorithm in which the sensors can identify their status in the distributed environment as "good" or "faulty"\newline & maximum accuracy and minimum false rate, complexity of the algorithm is low \\ 

Li et al. \cite{li2013research} & Perception Layer & addressed the challenges mentioned in "C2" by detecting threats involving the node security in IoT systems & improvement on the PKI-like security mechanism protocol\newline & improved security mechanism\\ 

Aggarwal et al. \cite{aggarwal2012rfid}  & Perception Layer & addressed the challenges mentioned in "C3", RFID security & proposed an improved protocol for Radio Frequency Identification (RFID) security & computationally efficient, prevent disclosure and desynchronization attacks.\\
	
Salami et al. \cite{al2016lightweight} & Perception Layer & addressed the challenges mentioned in "C4" speeding up the encryption operations & lightweight encryption scheme & more efficiency and reduced communication cost \\ 
	
Porambage et al. \cite{porambage2014pauthkey} & Perception Layer & addressed the challenges mentioned in "C5", resource constrained wireless sensor networks (WSNs) in distributed IoT application\newline & (PAuthKey) Pervasive authentication protocol and a key establishment scheme & end-users can authenticate themselves to the sensor nodes directly and acquire sensed data and services \\ 
\bottomrule
\end{tabular}
\end{table}
\end{center}

Salami et al. \cite{al2016lightweight} discussed confidentiality service, key management and efficiency of computation and communication issues by introducing a lightweight encryption scheme for smart homes. The scheme is effective for resource constrained devices and has flexible public key management. The result shows that the scheme is more efficient in the encryption operations and has reduced communication and computation overhead. Public key algorithm has high scalability and are considered favourable for node authentication without the need of complicated key management protocols \cite{miorandi2012internet}. 
Porambage et al. \cite{porambage2014pauthkey} proposed PAuthKey which is a pervasive lightweight authentication and key establishment scheme in distributed IoT application for the resource constrained wireless sensor networks (WSNs). PAuthKey scheme allows the end-users to establish secure connection, gain access to data and services and authenticate themselves directly with the sensor nodes. The security analysis and experimental results showed that the scheme is effective for the resource constrained WSN.

The authors \cite{chen2006distributed} \cite{li2013research} have proposed solutions for the threats involving node security in the perception layer. The author \cite{chen2006distributed} have proposed an algorithm to detect the faulty and compromised nodes while the author in \cite{li2013research} have proposed an improvement on the PKI-Like security meachanism. The author \cite{al2016lightweight} proposed an improved protocol for the RFID security that is embedded in the IoT devices. The author \cite{al2016lightweight} have proposed encryption and key management schemes to provide security solutions in the perception layer. The encryption mechanism is fast, efficient, reduced communication and computation overhead. The authors\cite{aggarwal2012rfid}\cite{al2016lightweight} have provided security solutions for resource-constrained device by proposing lightweight encryption techniques along with flexible public key management in \cite{aggarwal2012rfid} and lightweight authentication and key establishment schemes in \cite{al2016lightweight}. The pros and cons of the existing literature  that addressed security challenges, issues/attacks in perception layer are discussed in the Table \ref{perception}.

\subsection{Security solutions in the Network Layer}
Network layer is responsible for transmitting and processing the sensors data. Since it carries a large amount of data, it is highly prone to security attacks such as DoS, Man in the middle, DDoS attacks etc. Raza et al. \cite{raza2011securing} proposed a mechanism that supports End to End secure communication between the Internet and IP sensor networks. The proposed work features IPsec's Authentication Header (AH) and Encapsulation Security Payload (ESP) that enables the endpoints to authenticate, encrypt and check the integrity of the messages using traditional IPv6 mechanisms. Zhang et al. \cite{zhang2015communication} proposed a lightweight algorithm to prevent against DDoS attacks over IoT network environment. The author has tested the algorithm against different groups of network nodes such as working node, attacker node, monitoring node and legitimate user node. The author also describes that there's only one chance for an attacker's request to be served after that the packets will be dropped and the request is sent to the attacking list in the second attempt. The results showed that the proposed algorithm is capable of preventing and detecting Distributed Denial of Service attacks compared to the other existing systems. 

Salman et al.\cite{salman2016identity} discussed authentication techniques as one of the feature that would mitigate the security issues in IoT. They proposed identity-based authentication scheme to address the heterogeneity in IoT and to integrate the different protocols in IoT by applying the concept of Software Defined Networking (SDN) on IoT devices. The effectiveness of this scheme was tested using AVISPA tool and the evaluation showed that the scheme is immune to masquerade attack, man-in-the-middle attack, and replay attack. Santos et al. \cite{dos2015dtls} introduced a mutual authentication architecture that allows resource constrained devices to use Datagram Transport Layer Security (DTLS) for secure communication between the internet devices. The author also proposed a device called Internet of Things Security Support Provider (IoTSSP) which manages the certificates of the devices, provide authentication services and also responsible for establishing the session between the devices. The author also introduced two new main mechanisms i.e. Optional Handshaking Delegation and the Transfer of Session that prevents DoS attacks and also provide secured communication session. Similarly, Hummen et al.\cite{hummen2013towards} investigated the assumption of using certificates and lightweight security solutions for peer authentication in the IoT. The author also analysed the preliminary overhead reduction and discuss their applicability for the certificate-based DTLS handshake. The author designed three main ideas to reduce the overheads of the DTLS handshake which are based on session resumption, pre-validation and handshake delegation.

The authors \cite{zhang2015communication}\cite{dos2015dtls} have proposed different solutions against DoS, replay, man-in-the-middle and other attacks in the network layer. The author \cite{zhang2015communication} has tested the algorithm in different group of working nodes and are capable of detecting and preventing Distributed Denial of Service (DDoS) attacks whereas the author \cite{dos2015dtls} make use of mutual authentication architecture that allows resource constrained devices to use Datagram Transport Layer Security (DTLS) to prevent DoS attacks. The authors \cite{zhang2015communication}\cite{dos2015dtls} make use of DTLS to provide authentication service in the network layer. The author in \cite{salman2016identity} introduced two new main mechanisms i.e. Optional Handshaking Delegation and the Transfer of Session that prevents DoS attacks whereas the author \cite{hummen2013towards} designed three main ideas to reduce the overheads of the DTLS handshake which are based on session resumption, pre-validation and handshake delegation. The pros and cons of the existing literature  that addressed security challenges, issues/attacks in network layer are discussed in the Table \ref{network}.

\begin{center}
\begin{table}[h]
\centering
\caption{ Pros and cons of the addressed security challenges, issues/attacks in Network Layer of existing IoT systems\label{network}}%

\begin{tabular}{p{1.5cm} p{2cm} p{4cm} p{4cm} p{3cm} }  
\toprule
\textbf{Author and References} & \textbf{Addressed Layer} & \textbf{Issues Addressed}& \textbf{Proposed \newline Solutions}& \textbf{Pros and Cons of the existing solution} \\
\midrule
Raza et al. \cite{raza2011securing} & Network Layer & addressed the challenges mentioned in "C6", to authenticate, encrypt and check the integrity of the messages & a mechanism that supports End to End secure communication between the Internet and IP sensor networks & more secure and efficient \\ 

Zhang et al. \cite{zhang2015communication} & Network Layer & addressed the challenges mentioned in "C7", DDoS attacks & an algorithm to prevent DDoS attacks & Effective for detecting and preventing DDoS  \\

Salman et al.\cite{salman2016identity} & Network Layer & addressed the challenges mentioned in "C8", identity-based authentication & a novel IoT heterogeneous identity-based authentication scheme by applying the idea of Software Defined Networking (SDN) on IoT devices & resistant to man-in-the-middle attack, masquerade attack and replay attack  \\

Santos et al. \cite{dos2015dtls} & Network Layer & addressed the challenges mentioned in "C9", authentication problem to communicate with resource constrained device & architecture for secure communication between constrained IoT devices using Datagram Transport Layer Security (DTLS)  & prevents DoS attacks secure communication session  \\ 
    
Hummen et al.\cite{hummen2013towards} & Network Layer & addressed the challenges mentioned in "C10", peer authentication in the IoT & using certificates and lightweight security solutions\newline & overhead reduction \\ 

\bottomrule
\end{tabular}
\end{table}
\end{center}

\subsection{Security solutions in the Middleware Layer}
The middleware layer is responsible for information retrieval and processing and make decisions based on these processed results. Multi-party authentication, securely data storing in the cloud are some of the prime concerns in this layer. Tsai et al. \cite{tsai2015privacy} discussed the access control and authentication security concerns and proposed a user authentication technique over multiple servers. The communication and computational time between the multiple cloud service providers and traditional trusted third party service is reduced in the proposed scheme. The proposed scheme enables multiple cloud services from multiple service providers using one key and shows the scheme is efficient and secure. Shafagh et al. \cite{shafagh2015poster} introduced an Encrypted Query Processing approach that enable the system to store IoT data securely on the cloud database and allow query processing over the encrypted data. They make use of lightweight cryptographic algorithms for the resource constrained devices and the results showed that the system is efficient in database query processing and effective on low power and resource constrained devices. Horrow et al. \cite{horrow2012identity} proposed an identity management framework to authenticate data that are being transmitted between the cloud and the smart devices by placing an Identity Manager and Service Manager on the devices. The pros and cons of the existing literature  that addressed security challenges, issues/attacks in middleware ayer are discussed in the Table \ref{middle}.

\begin{center}
\begin{table}[h]
\centering
\caption{Pros and cons of the addressed security challenges, issues/attacks in Middleware Layer of existing IoT systems\label{middle}}%
 
\begin{tabular}{p{1.5cm} p{2cm} p{4cm} p{4cm} p{3cm} }  
\toprule
\textbf{Author and References} & \textbf{Addressed security solution Layers} & \textbf{Issues Addressed}& \textbf{Proposed Solutions}& \textbf{Pros and Cons of the existing solution} \\
\midrule
Tsai et al. \cite{tsai2015privacy} & Middleware Layer & addressed the challenges mentioned in "C11", Access and Authentication Control & Proposed a user authentication technique over multiple servers & Access multiple cloud services from multiple service providers using one key \\ 

Shafagh et al. \cite{shafagh2015poster} & Middleware Layer & addressed the challenges mentioned in "C12", store IoT data securely on the cloud database & introduced an Encrypted Query Processing approach to store IoT data securely on the cloud database and allow query processing over the encrypted data & efficient in database query processing and effective on low power and resource constrained devices \\ \\

Horrow et al. \cite{horrow2012identity} & Middleware Layer & addressed the challenges mentioned in "C13", to authenticate data that are being transmitted between the cloud and the smart devices  & by placing an Identity Manager and Service Manager on the devices & the protocols to develop the method have not yet been implemented \\
\bottomrule
\end{tabular}
\end{table}
\end{center}

\subsection{Security solutions in the Application Layer}
The application layer provide service to the end users. It is responsible for the messages exchange between application and end users. It is implemented using several protocols in the context of IoT. Some of the popular application layer protocols in IoT are Message Queuing Telemetry Transport (MQTT), Constrained Application Protocol (CoAP) and Extensible Messaging and Presence Protocol (XMPP) \cite{yassein2016application}. It includes smart devices that provide application specified services to the users. These smart device are generally simple, resource-constrained and light weight.that are vulnerable to the attacks.  
Seitz et al.\cite{seitz2013authorization} discussed access control and authorization in interconnected devices and proposed a framework that supports flexible access control and authorization for devices with low memory and processing power. The proposed framework provides remarkable flexibility for the access control models and minimize the communication cost when processing the exchanged message between the constrained and less constrained servers. Cirani et al. \cite{cirani2015iot} proposed IoT-OAS architecture to provide an authorization framework targeting HTTP/CoAP services, which can be integrated by invoking an external oauth-based authorization service (OAS). The proposed architecture is flexible and easy to integrate with external services and has the benefits in terms of low processing load, scalability and remote access customization. 

Park et al. \cite{park2015mutual} proposed a technique that used inter-device authentication and session-key distribution framework for secure communication between the devices. The proposed technique is capable of estimating the session key in prior that prevented attacks such as replay and man-in-the-middle attacks. Neisse et al. \cite{neisse2014enforcement} proposed a security policy to address the privacy and security challenges for the communication between the devices. The proposed policy provided optimal communication between the IoT devices. Tao et al. \cite{tao2010preference} proposed a mechanism for privacy protection that is preference-based for the IoT. Data and information privacy is one of the major concern in the application layer. The proposed mechanism introduces a trusted third party that evaluate the privacy preference of the users. The evaluated results and feedback are then sent to the service provider of the Internet of Things (SP). SP ensure desirable level of user privacy according to their preference and the third party provide supervision to the SP. The authors \cite{seitz2013authorization}\cite{cirani2015iot}\cite{park2015mutual} have proposed different frameworks as security solutions in the application layer. 

The author \cite{seitz2013authorization}\cite{cirani2015iot} have proposed access control models while in \cite{cirani2015iot} authorization framework targeting HTTP/CoAP services is proposed. The security of the network cannot alone guarantee the security attack in the IoT so the devices should also be manufactured with built-in security and patches. The security of IoT Systems can be further enhanced by applying Machine Learning / Artificial Intelligence Techniques and jamming techniques (Jamming the signal of the malicious nodes). The pros and cons of the existing literature  that addressed security challenges, issues/attacks in application layer are discussed in the Table \ref{application}. 

\begin{center}
\begin{table}[h]
\centering
\caption{Pros and cons of the addressed security challenges, issues/attacks in Application Layer of existing IoT systems\label{application}}%
 
\begin{tabular}{p{1.5cm} p{2cm} p{4cm} p{4cm} p{3cm} }  
\toprule
\textbf{Author and References} & \textbf{Addressed security solution Layers} & \textbf{Issues Addressed}& \textbf{Proposed Solutions}& \textbf{Pros and Cons of the existing solution} \\
\midrule
Seitz et al.\cite{seitz2013authorization} & Application Layer & addressed the challenges mentioned in "C14", access control and authorization issues in the resource constrained devices & proposed authorization framework, the decisions are based on local data and device's local conditions \newline & significant flexibility to the access control models \\ 

Cirani et al.\cite{cirani2015iot} & Application Layer & addressed the challenges mentioned in "C15", provide an authorization framework & an architecture IoT-OAS targeting HTTP/CoAP services & flexible, highly configurable, and easy to integrate with existing services,
lower processing load \newline \\ 

Park et al. \cite{park2015mutual} & Application Layer & addressed the challenges mentioned in "C16", to provide secure things-to-things communication & suggested an inter-device authentication and session key distribution system & Prevented replay attacks, man-inthe-middle attacks, estimated the session key in prior \\ \\ 

Neisse et al \cite{neisse2014enforcement} & Application Layer & addressed the challenges mentioned in "C17", privacy and security challenges for the communication between the devices & an enforcement security policy is suggested for addressing the privacy and security challenges & provided optimal communication
between the IoT devices \\ \\

Tao et al. \cite{tao2010preference} & Application Layer & addressed the challenges mentioned in "C18", privacy protection of the user & proposed a mechanism for privacy protection that is preference-based for the IoT & ensure desirable level of user privacy, not clear whether SP could perform supervision \\

\bottomrule
\end{tabular}
\end{table}
\end{center}

\section{Future research area}
 The development in the IoT systems will continue to evolve with more security and privacy challenges and to tackle these problems will always be the primary focus of the research in the advance IoT systems. It can be seen from table 2 that most of the security solutions for IoT are focused on authentication and authorization techniques. The security concern should not focus only on particular area or IoT layer but for the entire system. The proposed authentication and authorization techniques in the recent literature are still compromised for the resource constrained devices. There are more research to be done on the implementation of authentication techniques for resource constrained IoT devices that would be light-weight, energy-aware, fast and reliable. The investigation on Denial of Service attacks (DoS) that can manifest in IoT networks and a robust model that can quickly detect and eliminate such attacks is still in nascent stage. Generally, IoT devices are interconnected to the unreliable network via protocols like 6LoWPAN and IPv6. Despite of the good encryption and authentication mechanism, the devices might get exposed to the attacks from inside and outside the network. To counter these attacks, an intrusion detection technique could be useful. Abduvaliyev et. al \cite{abduvaliyev2013vital} surveyed the recent work done on intrusion detection systems and found most of the work proposed are for wireless sensor networks. The author presented a detailed classification of IDS techniques deployed in recent literature. The author highlighted the shortcomings of the currently employed intrusion detection systems and defined possible future attacks. The survey also gives an idea of future research directions that are still needed to fill the gaps. 
 
From our study, there are not many intrusion detection systems approach available for IPv6 connected IoT devices. The existing IDS approaches are available mainly for WSNs or traditional Internet. Raza et. al \cite{raza2013svelte} have designed an intrusion detection system called SVELTE that target attacks such as sinkhole, selective-forwarding, compromised information etc. The system is also capable of detecting malicious nodes and ring alarms during any detection. The design is suitable for the resource constrained devices. Smart phone users are continuously getting threatened by the malicious applications that get inside the device without any knowledge. These are generally termed as malware. The smart devices are highly prone to such malwares. Malwares are serious threats to user privacy and can damage the stored files and devices. Saracino et. al \cite{saracino2016madam} proposed a malware detection system called MADAM for android devices that are designed to detect any malicious behaviours. The system simultaneously analyses and block any malicious behaviours at four levels: kernel, suer, application and package. The author tested MADAM on three large datasets of 2800 apps and the proposed system efficiently blocks nearly 96\% of malicious apps with minimal power usage.

Overall, from our study we found security gaps in IoT architecture, lightweight security solutions and managing huge heterogeneous data. Security requirements for IoT are application specific. However, we can design a standard framework that can be customized according to the application requirements. We can apply software engineering concepts which can summarize the similarities of IoT applications and design a common framework that can provide the necessary security solutions to these applications.

Lightweight security solutions \cite{singh2017advanced} is an important area of future research direction due to resource constrained nature of IoT. Therefore, lightweight security concerns are noteworthy part that has to be well addressed before deploying advanced Internet of Things (IoT) systems. The security solutions such as key management, authentication, authorization, access control etc. should be energy aware and lightweight. The application computation and security requirements can be divided in different levels. Furthermore, we can add some default solutions to each levels so that it can satisfy the particular necessities of the levels. In this way the algorithm proposed in these levels can be modified to benefit the IoT system. Sometimes, remote authentication server might be required when there's unavailability of data because of the natural disaster and other factors. This has been a serious concern in the IoT system. Huang et al. \cite{huang2014robust} proposed two solutions to authenticate users from the remote locations. The first solution is a multi-factor authentication scheme that authenticates users by using passwords, bio-metrics and smart-cards. The other solution is a stand-alone authentication protocol which supports the authentication of the users even if the remote server is down. In comparison, the proposed techniques have shown efficient and reduced communication overhead. Similarly, ambient sensors and devices deployed in smart homes are vulnerable to many attacks. The IoT devices deployed at homes are generally pre-loaded with secret keys which the users usually do not update while installing at homes. Thus, it become easy for the attackers to hijack the wireless communications. The users might not be aware of this situation hence suffers from privacy issues. To address these problems Zhang et. al \cite{zhang2018matrix} have proposed a matrix based cross-layer key establishment protocol. The protocol enables smart home appliances to establish a secret session key among themselves without the use of pre-loaded secret keys. The protocol is promising in achieving secret session key with minimal energy consumption and without the need of pre-shared keys \cite{bashir2011energy}. Similarly some efficient authentication protocols~\cite{zeng2018aua,xu2018novel,xu2013algorithm} and abuse-free contract signing protocol with low-storage cost~\cite{xu2019csp} are proposed for various applications of IoT. Eventually, we may be ready to give a general unique IoT security system. 

IoT system produces huge amount of heterogeneous data each second. The efficient approach to manage these huge data produced by IoT systems is also focus of our research in future. We can apply technologies such as big data \cite{marjani2017big}, blockchain \cite{yu2018blockchain}, cloud and fog computing for the massive data exchange in a network. The technologies are efficient in managing heterogeneous huge data securely and efficiently. Fog computing \cite{alrawais2017fog} is emerged as a noteworthy technique that work closely to the edge of the network. The concept of fog computing is quite similar to cloud computing. It provides services such as data storage, massive data computing, application services etc. It has really re-defined the concept of cloud computing. Stojmenovic et al. \cite{stojmenovic2016overview} have discussed some techniques such as authentication and authorization as a security parameters in Fog computing. Atlam et al. \cite{atlam2018fog} explain what could be the security issues in the fog computing communications and lists some possible research directions to address these issues. A security scenario of fragile connection between cloud and fog computing is also addressed in this work with an example of authentication techniques. Therefore, it would be a good idea to apply big data, cloud and fog computing solutions on IoT systems to get comprehensive security solutions for the applications. 

Another interest research direction to look at is to combine formal methods with machine learning to detect possible security vulnerabilities in different layers of  IoT applications.  Formal method is able to provide a rigorous mathematical and logical guarantees for the safety and security properties of a given IoT application \cite{zheng2014state,zheng2015verification,zheng2017perceptions}. Though there are many restrictions in the approach (e.g., state space explosion, high cost of maintenance, and require non-trivial learning curve), recent work in the formal methods have improved to a practical level to address security issues both at development and run-time \cite{zheng2015braceassertion,zheng2014physically,zheng2017real,zheng2014braceassertion,zheng2013brace}.
Also machine learning has been increasingly used by both professionals and academics for various aspects in IoT \cite{zheng2018efficient, xie2018hybrid} and they can be used in combination with formal methods to provide rigorous yet scalable security guarantee for mission-critical IoT applications. 

\section{Conclusion}
The capacity and intelligence of IoT devices is wide open, like their exploitation.There is lack of standardization in the IoT market therefore every single connection could make the network vulnerable. A standard IoT architecture still remains an open issue. In this paper, we highlighted the security and privacy concerns in IoT domain. We have also reviewed the general architecture of IoT and examined the security issues at each layer in the IoT protocol stack i.e. perception layer, network layer and application layer. The paper also addressed the major challenges in securing IoT and discussed the security services requirements in IoT. Also, we have concisely presented an overview of existing approaches for securing the IoT systems. There's no consensus on the mechanism to apply security on IoT devices that are resource-constrained. The traditional network protocols and security mechanisms in IoT should be upgraded in order to suffice security requirements of this technology. Hence, security and privacy concerns are noteworthy part that has to be well studied before developing more advanced Internet of Things (IoT) systems.

\bibliography{wileyNJD-AMA}

@inproceedings{singh2014survey,
	title={A survey of Internet-of-Things: Future vision, architecture, challenges and services},
	author={Singh, Dhananjay and Tripathi, Gaurav and Jara, Antonio J},
	booktitle={Internet of things (WF-IoT), 2014 IEEE world forum on},
	pages={287--292},
	year={2014},
	organization={IEEE}
}

@inproceedings{khan2012future,
	title={Future internet: the internet of things architecture, possible applications and key challenges},
	author={Khan, Rafiullah and Khan, Sarmad Ullah and Zaheer, Rifaqat and Khan, Shahid},
	booktitle={Frontiers of Information Technology (FIT), 2012 10th International Conference on},
	pages={257--260},
	year={2012},
	organization={IEEE}
}

@article{atzori2010internet,
	title={The internet of things: A survey},
	author={Atzori, Luigi and Iera, Antonio and Morabito, Giacomo},
	journal={Computer networks},
	volume={54},
	number={15},
	pages={2787--2805},
	year={2010},
	publisher={Elsevier}
}

@inproceedings{zhao2013survey,
	title={A survey on the internet of things security},
	author={Zhao, Kai and Ge, Lina},
	booktitle={Computational Intelligence and Security (CIS), 2013 9th International Conference on},
	pages={663--667},
	year={2013},
	organization={IEEE}
}

@inproceedings{kanuparthi2013hardware,
	title={Hardware and embedded security in the context of internet of things},
	author={Kanuparthi, Arun and Karri, Ramesh and Addepalli, Sateesh},
	booktitle={Proceedings of the 2013 ACM workshop on Security, privacy \& dependability for cyber vehicles},
	pages={61--64},
	year={2013},
	organization={ACM}
}

@article{ Gartner,
	title={Gartner Press Release, online accesible at url={http://www.gartner.com/newsroom/id/2905717}}
    journal={Gartner},
	author={Gartner},
      url={http://www.gartner.com/newsroom/id/2905717}
}

@techreport{montenegro2007transmission,
  title={Transmission of IPv6 packets over IEEE 802.15. 4 networks},
  author={Montenegro, Gabriel and Kushalnagar, Nandakishore and Hui, Jonathan and Culler, David},
  year={2007}
}

@article{ieee2016p2413,
  title={P2413--Standard for an Architectural Framework for the Internet of Things (IoT)},
  author={IEEE Standards Association and others},
  journal={Institute of Electrical and Electronics Engineers, New York},
  year={2016}
}

@article{jing2014security,
	title={Security of the internet of things: Perspectives and challenges},
	author={Jing, Qi and Vasilakos, Athanasios V and Wan, Jiafu and Lu, Jingwei and Qiu, Dechao},
	journal={Wireless Networks},
	volume={20},
	number={8},
	pages={2481--2501},
	year={2014},
	publisher={Springer}
}

@article{internet2016international,
  title={International Telecommunication Union},
  author={Internet of Things Global Standards Initiative and others},
  journal={Retrieved Feb},
  volume={17},
  year={2016}
}

@article{adat2017security,
  title={Security in Internet of Things: issues, challenges, taxonomy, and architecture},
  author={Adat, Vipindev and Gupta, BB},
  journal={Telecommunication Systems},
  pages={1--19},
  year={2017},
  publisher={Springer}
}

@inproceedings{saadeh2016authentication,
	title={Authentication techniques for the internet of things: A survey},
	author={Saadeh, Maha and Sleit, Azzam and Qatawneh, Mohammed and Almobaideen, Wesam},
	booktitle={Cybersecurity and Cyberforensics Conference (CCC), 2016},
	pages={28--34},
	year={2016},
	organization={IEEE}
}

@inproceedings{kraijak2015survey,
	title={A survey on internet of things architecture, protocols, possible applications, security, privacy, real-world implementation and future trends},
	author={Kraijak, Surapon and Tuwanut, Panwit},
	booktitle={IEEE 16th International Conference on Communication Technology (ICCT)},
	pages={26--31},
	year={2015},
	organization={IEEE}
}

@inproceedings{kumar2016security,
	title={Security in internet of things: Challenges, solutions and future directions},
	author={Kumar, Sathish Alampalayam and Vealey, Tyler and Srivastava, Harshit},
	booktitle={System Sciences (HICSS), 2016 49th Hawaii International Conference on},
	pages={5772--5781},
	year={2016},
	organization={IEEE}
}

@article{turkanovic2014novel,
	title={A novel user authentication and key agreement scheme for heterogeneous ad hoc wireless sensor networks, based on the Internet of Things notion},
	author={Turkanovi{\'c}, Muhamed and Brumen, Bo{\v{s}}tjan and H{\"o}lbl, Marko},
	journal={Ad Hoc Networks},
	volume={20},
	pages={96--112},
	year={2014},
	publisher={Elsevier}
}

@article{padmavathi2009survey,
	title={A survey of attacks, security mechanisms and challenges in wireless sensor networks},
	author={Padmavathi, Dr G and Shanmugapriya, Mrs and others},
	journal={arXiv preprint arXiv:0909.0576},
	year={2009}
}

@article{padhy2011cloud,
	title={Cloud computing: security issues and research challenges},
	author={Padhy, Rabi Prasad and Patra, Manas Ranjan and Satapathy, Suresh Chandra},
	journal={International Journal of Computer Science and Information Technology \& Security (IJCSITS)},
	volume={1},
	number={2},
	pages={136--146},
	year={2011}
}

@article{fulare2011false,
	title={False data detection in wireless sensor network with secure communication},
	author={Fulare, Priyanka S and Chavhan, Nikita},
	journal={International Journal of Smart Sensors and AdHoc Networks (IJSSAN)},
	volume={1},
	year={2011}
}

@inproceedings{talwana2016smart,
	title={Smart World of Internet of Things (IoT) and Its Security Concerns},
	author={Talwana, Jonathan Charity and Hua, Huang Jian},
	booktitle={Internet of Things (iThings) and IEEE Green Computing and Communications (GreenCom) and IEEE Cyber, Physical and Social Computing (CPSCom) and IEEE Smart Data (SmartData), 2016 IEEE International Conference on},
	pages={240--245},
	year={2016},
	organization={IEEE}
}

@article{weber2015internet,
	title={Internet of things: Privacy issues revisited},
	author={Weber, Rolf H},
	journal={Computer Law \& Security Review},
	volume={31},
	number={5},
	pages={618--627},
	year={2015},
	publisher={Elsevier}
}

@inproceedings{vasilomanolakis2015security,
	title={On the Security and Privacy of Internet of Things Architectures and Systems},
	author={Vasilomanolakis, Emmanouil and Daubert, J{\"o}rg and Luthra, Manisha and Gazis, Vangelis and Wiesmaier, Alex and Kikiras, Panayotis},
	booktitle={Secure Internet of Things (SIoT), 2015 International Workshop on},
	pages={49--57},
	year={2015},
	organization={IEEE}
}

@article{miorandi2012internet,
	title={Internet of things: Vision, applications and research challenges},
	author={Miorandi, Daniele and Sicari, Sabrina and De Pellegrini, Francesco and Chlamtac, Imrich},
	journal={Ad Hoc Networks},
	volume={10},
	number={7},
	pages={1497--1516},
	year={2012},
	publisher={Elsevier}
}

@article{bhabad2015internet,
	title={Internet of things: architecture, security issues and countermeasures},
	author={Bhabad, Mayuri A and Bagade, Sudhir T},
	journal={International Journal of Computer Applications},
	volume={125},
	number={14},
	year={2015},
	publisher={Foundation of Computer Science}
}

@inproceedings{chen2006distributed,
	title={Distributed fault detection of wireless sensor networks},
	author={Chen, Jinran and Kher, Shubha and Somani, Arun},
	booktitle={Proceedings of the 2006 workshop on Dependability issues in wireless ad hoc networks and sensor networks},
	pages={65--72},
	year={2006},
	organization={ACM}
}

@inproceedings{li2013research,
	title={Research on PKI-like Protocol for the Internet of Things},
	author={Li, Zhihua and Yin, Xi and Geng, Zhenmin and Zhang, Haitao and Li, Pengfei and Sun, Ya and Zhang, Huawei and Li, Lin},
	booktitle={Measuring Technology and Mechatronics Automation (ICMTMA), 2013 Fifth International Conference on},
	pages={915--918},
	year={2013},
	organization={IEEE}
}

@inproceedings{aggarwal2012rfid,
  title={RFID security in the context of internet of things},
  author={Aggarwal, Renu and Das, Manik Lal},
  booktitle={Proceedings of the First International Conference on Security of Internet of Things},
  pages={51--56},
  year={2012},
  organization={ACM}
}

@inproceedings{al2016lightweight,
	title={Lightweight encryption for smart home},
	author={Al Salami, Sanaah and Baek, Joonsang and Salah, Khaled and Damiani, Ernesto},
	booktitle={Availability, Reliability and Security (ARES), 2016 11th International Conference on},
	pages={382--388},
	year={2016},
	organization={IEEE}
}

@article{porambage2014pauthkey,
	title={PAuthKey: A pervasive authentication protocol and key establishment scheme for wireless sensor networks in distributed IoT applications},
	author={Porambage, Pawani and Schmitt, Corinna and Kumar, Pardeep and Gurtov, Andrei and Ylianttila, Mika},
	journal={International Journal of Distributed Sensor Networks},
	volume={10},
	number={7},
	pages={357430},
	year={2014},
	publisher={SAGE Publications Sage UK: London, England}
}

@inproceedings{raza2011securing,
	title={Securing communication in 6LoWPAN with compressed IPsec},
	author={Raza, Shahid and Duquennoy, Simon and Chung, Tony and Yazar, Dogan and Voigt, Thiemo and Roedig, Utz},
	booktitle={Distributed Computing in Sensor Systems and Workshops (DCOSS), 2011 International Conference on},
	pages={1--8},
	year={2011},
	organization={IEEE}
}

@inproceedings{zhang2015communication,
	title={Communication security in internet of thing: preventive measure and avoid DDoS attack over IoT network},
	author={Zhang, Congyingzi and Green, Robert},
	booktitle={Proceedings of the 18th Symposium on Communications \& Networking},
	pages={8--15},
	year={2015},
	organization={Society for Computer Simulation International}
}

@inproceedings{salman2016identity,
	title={Identity-based authentication scheme for the internet of things},
	author={Salman, Ola and Abdallah, Sarah and Elhajj, Imad H and Chehab, Ali and Kayssi, Ayman},
	booktitle={Computers and Communication (ISCC), 2016 IEEE Symposium on},
	pages={1109--1111},
	year={2016},
	organization={IEEE}
}

@inproceedings{dos2015dtls,
	title={A DTLS-based security architecture for the Internet of Things},
	author={dos Santos, Giederson Lessa and Guimaraes, Vinicius Tavares and da Cunha Rodrigues, Guilherme and Granville, Lisandro Zambenedetti and Tarouco, Liane Margarida Rockenbach},
	booktitle={Computers and Communication (ISCC), 2015 IEEE Symposium on},
	pages={809--815},
	year={2015},
	organization={IEEE}
}

@inproceedings{hummen2013towards,
	title={Towards viable certificate-based authentication for the internet of things},
	author={Hummen, Ren{\'e} and Ziegeldorf, Jan H and Shafagh, Hossein and Raza, Shahid and Wehrle, Klaus},
	booktitle={Proceedings of the 2nd ACM workshop on Hot topics on wireless network security and privacy},
	pages={37--42},
	year={2013},
	organization={ACM}
}

@article{tsai2015privacy,
	title={A privacy-aware authentication scheme for distributed mobile cloud computing services},
	author={Tsai, Jia-Lun and Lo, Nai-Wei},
	journal={IEEE systems journal},
	volume={9},
	number={3},
	pages={805--815},
	year={2015},
	publisher={IEEE}
}

@inproceedings{shafagh2015poster,
	title={Poster: Towards encrypted query processing for the Internet of Things},
	author={Shafagh, Hossein and Hithnawi, Anwar and Dr{\"o}scher, Andreas and Duquennoy, Simon and Hu, Wen},
	booktitle={Proceedings of the 21st Annual International Conference on Mobile Computing and Networking},
	pages={251--253},
	year={2015},
	organization={ACM}
}

@inproceedings{horrow2012identity,
  title={Identity management framework for cloud based internet of things},
  author={Horrow, Susmita and Sardana, Anjali},
  booktitle={Proceedings of the First International Conference on Security of Internet of Things},
  pages={200--203},
  year={2012},
  organization={ACM}
}

@inproceedings{seitz2013authorization,
	title={Authorization framework for the internet-of-things},
	author={Seitz, Ludwig and Selander, G{\"o}ran and Gehrmann, Christian},
	booktitle={World of Wireless, Mobile and Multimedia Networks (WoWMoM), 2013 IEEE 14th International Symposium and Workshops on a},
	pages={1--6},
	year={2013},
	organization={IEEE}
}

@article{cirani2015iot,
	title={Iot-oas: An oauth-based authorization service architecture for secure services in iot scenarios},
	author={Cirani, Simone and Picone, Marco and Gonizzi, Pietro and Veltri, Luca and Ferrari, Gianluigi},
	journal={IEEE sensors journal},
	volume={15},
	number={2},
	pages={1224--1234},
	year={2015},
	publisher={IEEE}
}

@article{park2015mutual,
  title={Mutual authentication scheme in secure internet of things technology for comfortable lifestyle},
  author={Park, Namje and Kang, Namhi},
  journal={Sensors},
  volume={16},
  number={1},
  pages={20},
  year={2015},
  publisher={Multidisciplinary Digital Publishing Institute}
}

@inproceedings{neisse2014enforcement,
  title={Enforcement of security policy rules for the internet of things},
  author={Neisse, Ricardo and Steri, Gary and Baldini, Gianmarco},
  booktitle={Wireless and Mobile Computing, Networking and Communications (WiMob), 2014 IEEE 10th International Conference on},
  pages={165--172},
  year={2014},
  organization={IEEE}
}

@inproceedings{tao2010preference,
  title={Preference-based privacy protection mechanism for the internet of things},
  author={Tao, Hu and Peiran, Wang},
  booktitle={Information Science and Engineering (ISISE), 2010 International Symposium on},
  pages={531--534},
  year={2010},
  organization={IEEE}
}

@article{abduvaliyev2013vital,
  title={On the vital areas of intrusion detection systems in wireless sensor networks},
  author={Abduvaliyev, Abror and Pathan, Al-Sakib Khan and Zhou, Jianying and Roman, Rodrigo and Wong, Wai-Choong},
  journal={IEEE Communications Surveys \& Tutorials},
  volume={15},
  number={3},
  pages={1223--1237},
  year={2013},
  publisher={IEEE}
}

@article{raza2013svelte,
  title={SVELTE: Real-time intrusion detection in the Internet of Things},
  author={Raza, Shahid and Wallgren, Linus and Voigt, Thiemo},
  journal={Ad hoc networks},
  volume={11},
  number={8},
  pages={2661--2674},
  year={2013},
  publisher={Elsevier}
}

@article{saracino2016madam,
  title={Madam: Effective and efficient behavior-based android malware detection and prevention},
  author={Saracino, Andrea and Sgandurra, Daniele and Dini, Gianluca and Martinelli, Fabio},
  journal={IEEE Transactions on Dependable and Secure Computing},
  year={2016},
  publisher={IEEE}
}

@article{huang2014robust,
  title={Robust multi-factor authentication for fragile communications},
  author={Huang, Xinyi and Xiang, Yang and Bertino, Elisa and Zhou, Jianying and Xu, Li},
  journal={IEEE Transactions on Dependable and Secure Computing},
  volume={11},
  number={6},
  pages={568--581},
  year={2014},
  publisher={IEEE}
}

@article{zhang2018matrix,
  title={A matrix-based cross-layer key establishment protocol for smart homes},
  author={Zhang, Yuexin and Xiang, Yang and Huang, Xinyi and Chen, Xiaofeng and Alelaiwi, Abdulhameed},
  journal={Information Sciences},
  volume={429},
  pages={390--405},
  year={2018},
  publisher={Elsevier}
}

@article{stojmenovic2016overview,
  title={An overview of fog computing and its security issues},
  author={Stojmenovic, Ivan and Wen, Sheng and Huang, Xinyi and Luan, Hao},
  journal={Concurrency and Computation: Practice and Experience},
  volume={28},
  number={10},
  pages={2991--3005},
  year={2016},
  publisher={Wiley Online Library}
}

@inproceedings{suo2012security,
  title={Security in the internet of things: a review},
  author={Suo, Hui and Wan, Jiafu and Zou, Caifeng and Liu, Jianqi},
  booktitle={2012 international conference on computer science and electronics engineering},
  pages={648--651},
  year={2012},
  organization={IEEE}
}

@article{zhang2019driver,
  title={Driver Drowsiness Detection using Multi-Channel Second Order Blind Identifications},
  author={Zhang, Chao and Wu, Xiaopei and Zheng, Xi and Yu, Shui},
  journal={IEEE Access},
  year={2019},
  publisher={IEEE}
}

@inproceedings{bhandari2017non,
  title={Non-invasive sensor based automated smoking activity detection},
  author={Bhandari, Babin and Lu, JianChao and Zheng, Xi and Rajasegarar, Sutharshan and Karmakar, Chandan},
  booktitle={2017 39th Annual International Conference of the IEEE Engineering in Medicine and Biology Society (EMBC)},
  pages={845--848},
  year={2017},
  organization={IEEE}
}

@inproceedings{abkenar2017service,
  title={Service-Mediated On-Road Situation-Awareness for Group Activity Safety},
  author={Abkenar, Amin B and Loke, Seng W and Zheng, James Xi and Zaslavsky, Arkady},
  booktitle={Proceedings of the 14th EAI International Conference on Mobile and Ubiquitous Systems: Computing, Networking and Services},
  pages={478--481},
  year={2017},
  organization={ACM}
}

@inproceedings{lu2019detection,
  title={Detection of Smoking Events from Confounding Activities of Daily Living},
  author={Lu, Jianchao and Wang, Jiaxing and Zheng, Xi and Karmakar, Chandan and Rajasegarar, Sutharshan},
  booktitle={Proceedings of the Australasian Computer Science Week Multiconference},
  pages={39},
  year={2019},
  organization={ACM}
}

@article{yu2018survey,
  title={A survey on security issues in services communication of Microservices-enabled fog applications},
  author={Yu, Dongjin and Jin, Yike and Zhang, Yuqun and Zheng, Xi},
  journal={Concurrency and Computation: Practice and Experience},
  pages={e4436},
  year={2018},
  publisher={Wiley Online Library}
}

@article{pan2017cyber,
  title={Cyber security attacks to modern vehicular systems},
  author={Pan, Lei and Zheng, Xi and Chen, HX and Luan, T and Bootwala, Huzefa and Batten, Lynn},
  journal={Journal of information security and applications},
  volume={36},
  pages={90--100},
  year={2017},
  publisher={Elsevier}
}

@inproceedings{zheng2017security,
  title={Security analysis of modern mission critical android mobile applications},
  author={Zheng, Xi and Pan, Lei and Yilmaz, Erdem},
  booktitle={Proceedings of the Australasian Computer Science Week Multiconference},
  pages={2},
  year={2017},
  organization={ACM}
}

@article{radhappa2018practical,
  title={Practical overview of security issues in wireless sensor network applications},
  author={Radhappa, Harish and Pan, Lei and Xi Zheng, James and Wen, Sheng},
  journal={International journal of computers and applications},
  volume={40},
  number={4},
  pages={202--213},
  year={2018},
  publisher={Taylor \& Francis}
}

@inproceedings{zheng2017testbed,
  title={A testbed for security analysis of modern vehicle systems},
  author={Zheng, Xi and Pan, Lei and Chen, Hongxu and Di Pietro, Rick and Batten, Lynn},
  booktitle={2017 IEEE Trustcom/BigDataSE/ICESS},
  pages={1090--1095},
  year={2017},
  organization={IEEE}
}

@inproceedings{zheng2016investigating,
  title={Investigating security vulnerabilities in modern vehicle systems},
  author={Zheng, Xi and Pan, Lei and Chen, Hongxu and Wang, Peiyin},
  booktitle={International Conference on Applications and Techniques in Information Security},
  pages={29--40},
  year={2016},
  organization={Springer}
}

@article{zeng2018aua,
  title={E-AUA: An Efficient Anonymous User Authentication Protocol for Mobile IoT},
  author={Zeng, Xianjiao and Xu, Guangquan and Zheng, Xi and Xiang, Yang and Zhou, Wanlei},
  journal={IEEE Internet of Things Journal},
  year={2018},
  publisher={IEEE}
}

@article{xu2019csp,
  title={CSP-E2: An abuse-free contract signing protocol with low-storage TTP for energy-efficient electronic transaction ecosystems},
  author={Xu, Guangquan and Zhang, Yao and Sangaiah, Arun Kumar and Li, Xiaohong and Castiglione, Aniello and Zheng, Xi},
  journal={Information Sciences},
  volume={476},
  pages={505--515},
  year={2019},
  publisher={Elsevier}
}

@article{zheng2014state,
  title={On the state of the art in verification and validation in cyber physical systems},
  author={Zheng, Xi and Julien, Christine and Kim, Miryung and Khurshid, Sarfraz},
  journal={The University of Texas at Austin, The Center for Advanced Research in Software Engineering, Tech. Rep. TR-ARiSE-2014-001},
  volume={1485},
  year={2014},
  publisher={Citeseer}
}

@inproceedings{zheng2015verification,
  title={Verification and validation in cyber physical systems: research challenges and a way forward},
  author={Zheng, Xi and Julien, Christine},
  booktitle={2015 IEEE/ACM 1st International Workshop on Software Engineering for Smart Cyber-Physical Systems},
  pages={15--18},
  year={2015},
  organization={IEEE}
}

@article{zheng2017perceptions,
  title={Perceptions on the state of the art in verification and validation in cyber-physical systems},
  author={Zheng, Xi and Julien, Christine and Kim, Miryung and Khurshid, Sarfraz},
  journal={IEEE Systems Journal},
  volume={11},
  number={4},
  pages={2614--2627},
  year={2017},
  publisher={IEEE}
}

@article{zheng2018efficient,
  title={Efficient and scalable runtime monitoring for cyber--physical system},
  author={Zheng, Xi and Julien, Christine and Podorozhny, Rodion and Cassez, Franck and Rakotoarivelo, Thierry},
  journal={IEEE Systems Journal},
  volume={12},
  number={2},
  pages={1667--1678},
  year={2018},
  publisher={IEEE}
}

@inproceedings{zheng2015braceassertion,
  title={Braceassertion: Runtime verification of cyber-physical systems},
  author={Zheng, Xi and Julien, Christine and Podorozhny, Rodion and Cassez, Franck},
  booktitle={2015 IEEE 12th International Conference on Mobile Ad Hoc and Sensor Systems},
  pages={298--306},
  year={2015},
  organization={IEEE}
}

@inproceedings{zheng2014physically,
  title={Physically informed assertions for cyber physical systems development and debugging},
  author={Zheng, Xi},
  booktitle={2014 IEEE International Conference on Pervasive Computing and Communication Workshops (PERCOM WORKSHOPS)},
  pages={181--183},
  year={2014},
  organization={IEEE}
}

@article{zheng2017real,
  title={Real-time simulation support for runtime verification of cyber-physical systems},
  author={Zheng, Xi and Julien, Christine and Chen, Hongxu and Podorozhny, Rodion and Cassez, Franck},
  journal={ACM Transactions on Embedded Computing Systems (TECS)},
  volume={16},
  number={4},
  pages={106},
  year={2017},
  publisher={ACM}
}

@misc{zheng2014braceassertion,
  title={Braceassertion: Behavior-driven development for cps application},
  author={Zheng, Xi and Julien, Christine and Podorozhny, Rodion and Cassez, Franck},
  year={2014},
  publisher={Technical Report UTARISE-2015-002}
}

@article{zheng2013brace,
  title={Brace: Assertion-driven development of cyber-physical systems applications},
  author={Zheng, Xi and Fok, Chien-Liang and Julien, Christine and Khurshid, Sarfraz and Kim, Miryung},
  journal={Tech. Report TR-ARiSE-2013-001, University of Texas at Austin},
  year={2013}
}

@article{xie2018hybrid,
  title={A Hybrid Method Combining Markov Prediction and Fuzzy Classification for Driving Condition Recognition},
  author={Xie, Haiming and Tian, Guangyu and Du, Guangqian and Huang, Yong and Chen, Hongxu and Zheng, Xi and Luan, Tom H},
  journal={IEEE Transactions on Vehicular Technology},
  volume={67},
  number={11},
  pages={10411--10424},
  year={2018},
  publisher={IEEE}
}

@inproceedings{zhang2018soprotector,
  title={SoProtector: Securing Native C/C++ Libraries for Mobile Applications},
  author={Zhang, Ning and Xu, Guangquan and Meng, Guozhu and Zheng, Xi},
  booktitle={International Conference on Algorithms and Architectures for Parallel Processing},
  pages={417--431},
  year={2018},
  organization={Springer}
}

@article{xu2018novel,
  title={A novel efficient MAKA protocol with desynchronization for anonymous roaming service in Global Mobility Networks},
  author={Xu, Guangquan and Liu, Jia and Lu, Yanrong and Zeng, Xianjiao and Zhang, Yao and Li, Xiaoming},
  journal={Journal of Network and Computer Applications},
  volume={107},
  pages={83--92},
  year={2018},
  publisher={Elsevier}
}

@article{xu2013algorithm,
  title={An algorithm on fairness verification of mobile sink routing in wireless sensor network},
  author={Xu, Guangquan and Li, Weisheng and Xu, Rui and Xiao, Yingyuan and Gao, Honghao and Li, Xiaohong and Feng, Zhiyong and Mei, Jia},
  journal={Personal and ubiquitous computing},
  volume={17},
  number={5},
  pages={851--864},
  year={2013},
  publisher={Springer}
}

@inproceedings{yassein2016application,
  title={Application layer protocols for the Internet of Things: A survey},
  author={Yassein, Muneer Bani and Shatnawi, Mohammed Q and others},
  booktitle={2016 International Conference on Engineering \& MIS (ICEMIS)},
  pages={1--4},
  year={2016},
  organization={IEEE}
}

@article{panarello2018blockchain,
  title={Blockchain and IoT integration: A systematic survey},
  author={Panarello, Alfonso and Tapas, Nachiket and Merlino, Giovanni and Longo, Francesco and Puliafito, Antonio},
  journal={Sensors},
  volume={18},
  number={8},
  pages={2575},
  year={2018},
  publisher={Multidisciplinary Digital Publishing Institute}
}

@article{khan2018iot,
  title={IoT security: Review, blockchain solutions, and open challenges},
  author={Khan, Minhaj Ahmad and Salah, Khaled},
  journal={Future Generation Computer Systems},
  volume={82},
  pages={395--411},
  year={2018},
  publisher={Elsevier}
}

@article{vaughan2003mobile,
  title={Mobile IPv6 and the future of wireless Internet access},
  author={Vaughan-Nichols, Steven J},
  journal={Computer},
  volume={36},
  number={2},
  pages={18--20},
  year={2003},
  publisher={IEEE}
}

@incollection{kumar2018review,
  title={A Review of Low-Power VLSI Technology Developments},
  author={Kumar, Nakka Ravi},
  booktitle={Innovations in Electronics and Communication Engineering},
  pages={17--27},
  year={2018},
  publisher={Springer}
}

@article{silva2018internet,
  title={Internet of things: A comprehensive review of enabling technologies, architecture, and challenges},
  author={Silva, Bhagya Nathali and Khan, Murad and Han, Kijun},
  journal={IETE Technical review},
  volume={35},
  number={2},
  pages={205--220},
  year={2018},
  publisher={Taylor \& Francis}
}

@article{puthal2016threats,
  title={Threats to networking cloud and edge datacenters in the Internet of Things},
  author={Puthal, Deepak and Nepal, Surya and Ranjan, Rajiv and Chen, Jinjun},
  journal={IEEE Cloud Computing},
  volume={3},
  number={3},
  pages={64--71},
  year={2016},
  publisher={IEEE}
}

@article{bello2017network,
  title={Network layer inter-operation of Device-to-Device communication technologies in Internet of Things (IoT)},
  author={Bello, Oladayo and Zeadally, Sherali and Badra, Mohamad},
  journal={Ad Hoc Networks},
  volume={57},
  pages={52--62},
  year={2017},
  publisher={Elsevier}
}

@article{da2018reference,
  title={A reference model for internet of things middleware},
  author={da Cruz, Mauro AA and Rodrigues, Joel Jos{\'e} PC and Al-Muhtadi, Jalal and Korotaev, Valery V and de Albuquerque, Victor Hugo C},
  journal={IEEE Internet of Things Journal},
  volume={5},
  number={2},
  pages={871--883},
  year={2018},
  publisher={IEEE}
}

@article{raja2018internet,
  title={Internet of things: Challenges, issues and applications},
  author={Raja, SP and Rajkumar, T Dhiliphan and Raj, Vivek Pandiya},
  journal={Journal of Circuits, Systems and Computers},
  volume={27},
  number={12},
  pages={1830007},
  year={2018},
  publisher={World Scientific}
}

@article{singh2017advanced,
  title={Advanced lightweight encryption algorithms for IoT devices: survey, challenges and solutions},
  author={Singh, Saurabh and Sharma, Pradip Kumar and Moon, Seo Yeon and Park, Jong Hyuk},
  journal={Journal of Ambient Intelligence and Humanized Computing},
  pages={1--18},
  year={2017},
  publisher={Springer}
}

@article{bouloukakis2019automated,
  title={Automated synthesis of mediators for middleware-layer protocol interoperability in the IoT},
  author={Bouloukakis, Georgios and Georgantas, Nikolaos and Ntumba, Patient and Issarny, Val{\'e}rie},
  journal={Future Generation Computer Systems},
  year={2019},
  publisher={Elsevier}
}

@inproceedings{razouk2017new,
  title={A new security middleware architecture based on fog computing and cloud to support IoT constrained devices},
  author={Razouk, Wissam and Sgandurra, Daniele and Sakurai, Kouichi},
  booktitle={Proceedings of the 1st International Conference on Internet of Things and Machine Learning},
  pages={35},
  year={2017},
  organization={ACM}
}

@article{sethi2017internet,
  title={Internet of things: architectures, protocols, and applications},
  author={Sethi, Pallavi and Sarangi, Smruti R},
  journal={Journal of Electrical and Computer Engineering},
  volume={2017},
  year={2017},
  publisher={Hindawi}
}

@article{alrawais2017fog,
  title={Fog computing for the internet of things: Security and privacy issues},
  author={Alrawais, Arwa and Alhothaily, Abdulrahman and Hu, Chunqiang and Cheng, Xiuzhen},
  journal={IEEE Internet Computing},
  volume={21},
  number={2},
  pages={34--42},
  year={2017},
  publisher={IEEE}
}

@article{marjani2017big,
  title={Big IoT data analytics: architecture, opportunities, and open research challenges},
  author={Marjani, Mohsen and Nasaruddin, Fariza and Gani, Abdullah and Karim, Ahmad and Hashem, Ibrahim Abaker Targio and Siddiqa, Aisha and Yaqoob, Ibrar},
  journal={IEEE Access},
  volume={5},
  pages={5247--5261},
  year={2017},
  publisher={IEEE}
}

@article{atlam2018fog,
  title={Fog computing and the internet of things: a review},
  author={Atlam, Hany and Walters, Robert and Wills, Gary},
  journal={big data and cognitive computing},
  volume={2},
  number={2},
  pages={10},
  year={2018},
  publisher={Multidisciplinary Digital Publishing Institute}
}

@incollection{serbanati2011building,
  title={Building blocks of the internet of things: State of the art and beyond},
  author={Serbanati, Alexandru and Medaglia, Carlo Maria and Ceipidor, Ugo Biader},
  booktitle={Deploying RFID-Challenges, Solutions, and Open Issues},
  year={2011},
  publisher={IntechOpen}
}

@inproceedings{jolfaei2019privacy,
  title={Privacy and Security of Connected Vehicles in Intelligent Transportation System},
  author={Jolfaei, Alireza and Kant, Krishna},
  booktitle={2019 49th Annual IEEE/IFIP International Conference on Dependable Systems and Networks},
  pages={9--10},
  year={2019},
  organization={IEEE}
}

@article{jolfaei2015preserving,
  title={Preserving the confidentiality of digital images using a chaotic encryption scheme.},
  author={Jolfaei, Alireza and Matinfar, Ahmadreza and Mirghadri, Abdolrasoul},
  journal={IJESDF},
  volume={7},
  number={3},
  pages={258--277},
  year={2015}
}

@article{jolfaei2011substitution,
  title={Substitution-permutation based image cipher using chaotic Henon and Baker’s maps},
  author={Jolfaei, Alireza and Mirghadri, Abdolrasoul},
  journal={International Review on Computers and Software},
  volume={6},
  number={1},
  pages={40--54},
  year={2011},
  publisher={Praise Worthy Prize}
}

@inproceedings{jolfaei2015secure,
  title={A secure lightweight texture encryption scheme},
  author={Jolfaei, Alireza and Wu, Xin-Wen and Muthukkumarasamy, Vallipuram},
  booktitle={Image and Video Technology},
  pages={344--356},
  year={2015},
  organization={Springer}
}

@article{jolfaei2012impact,
  title={Impact of Rotations in the Salsa 20/8 Image Encryption Scheme},
  author={Jolfaei, Alireza and Mirghadri, Abdolrasoul and Vizandan, Ahmadreza},
  journal={International Journal of Computer Theory and Engineering},
  volume={4},
  number={6},
  pages={938},
  year={2012},
  publisher={IACSIT Press}
}

@inproceedings{jolfaei2017lightweight,
  title={A Lightweight Integrity Protection Scheme for Fast Communications in Smart Grid.},
  author={Jolfaei, Alireza and Kant, Krishna},
  booktitle={SECRYPT},
  pages={31--42},
  year={2017}
}

@article{abbasi2019generalized,
  title={Generalized PVO-based dynamic block reversible data hiding for secure transmission using firefly algorithm},
  author={Abbasi, Rashid and Faseeh Qureshi, Nawab Muhammad and Hassan, Haseeb and Saba, Tanzila and Rehman, Amjad and Luo, Bin and Bashir, Ali Kashif},
  journal={Transactions on Emerging Telecommunications Technologies},
  pages={e3680},
  year={2019},
  publisher={Wiley Online Library}
}

@article{musaddiq2018survey,
  title={A survey on resource management in IoT operating systems},
  author={Musaddiq, Arslan and Zikria, Yousaf Bin and Hahm, Oliver and Yu, Heejung and Bashir, Ali Kashif and Kim, Sung Won},
  journal={IEEE Access},
  volume={6},
  pages={8459--8482},
  year={2018},
  publisher={IEEE}
}

@article{bashir2011energy,
  title={Energy efficient in-network RFID data filtering scheme in wireless sensor networks},
  author={Bashir, Ali Kashif and Lim, Se-Jung and Hussain, Chauhdary Sajjad and Park, Myong-Soon},
  journal={Sensors},
  volume={11},
  number={7},
  pages={7004--7021},
  year={2011},
  publisher={Molecular Diversity Preservation International}
}

@article{ali2018quality,
  title={Quality of Service Provisioning for Heterogeneous Services in Cognitive Radio-enabled Internet of Things},
  author={Ali, Amjad and Feng, Li and Bashir, Ali Kashif and El-Sappagh, Shaker Hassan A and Ahmed, Syed Hassan and Iqbal, Muddesar and Raja, Gunsekaran},
  journal={IEEE Transactions on Network Science and Engineering},
  year={2018},
  publisher={IEEE}
}

@article{lin2017survey,
  title={A survey on internet of things: Architecture, enabling technologies, security and privacy, and applications},
  author={Lin, Jie and Yu, Wei and Zhang, Nan and Yang, Xinyu and Zhang, Hanlin and Zhao, Wei},
  journal={IEEE Internet of Things Journal},
  volume={4},
  number={5},
  pages={1125--1142},
  year={2017},
  publisher={IEEE}
}

@article{yu2018blockchain,
  title={Blockchain-based solutions to security and privacy issues in the internet of things},
  author={Yu, Yong and Li, Yannan and Tian, Junfeng and Liu, Jianwei},
  journal={IEEE Wireless Communications},
  volume={25},
  number={6},
  pages={12--18},
  year={2018},
  publisher={IEEE}
}

\end{document}